\documentclass[english]{llncs}
\usepackage[T1]{fontenc}
\usepackage[latin9]{inputenc}
\usepackage{array}
\usepackage{verbatim}
\usepackage{url}
\usepackage{amstext}
\usepackage{graphicx}

\makeatletter

\providecommand{\tabularnewline}{\\}

\usepackage{url}

\@ifundefined{showcaptionsetup}{}{%
 \PassOptionsToPackage{caption=false}{subfig}}
\usepackage{subfig}
\makeatother

\usepackage{babel}

\begin{document}

\title{A Comparative Case Study of Code Reuse With\\ Language Oriented Programming\thanks{\small This research was supported in part by the \emph{Israel Science Foundation (ISF)} under grant No. 926/08.}}

\author{{David H. Lorenz\inst{1}\and Boaz Rosenan\inst{2}\\[0.1in] \small Open University of Israel,\\ 1 University Rd., P.O.Box 808, Raanana 43107 Israel,}\institute{\email{lorenz@openu.ac.il},\and \email{brosenan@cslab.openu.ac.il}}}
\maketitle
\begin{abstract}
There is a gap between our ability to reuse high-level concepts in
software design and our ability to reuse the code implementing them.
%
{}Language Oriented Programming (LOP) is a software development paradigm
that aims to close this gap, through extensive use of Domain Specific
Languages (DSLs).  With LOP, the high-level reusable concepts become
reusable DSL constructs, and their translation into code level concepts
is done in the DSL implementation. Particular products are implemented
using DSL code, thus reusing only high-level concepts.  In this paper
we provide a comparison between two implementation approaches for
LOP: (a)\,using external DSLs with a projectional language workbench
(MPS); and (b)\,using internal DSLs with an LOP language (Cedalion).
 To demonstrate how reuse is achieved in each approach, we present
a small case study, where LOP is used to build a Software Product
Line (SPL) of calculator software.
\end{abstract}

\section{Introduction}

A key issue with software reuse is the gap between concept reuse
and code reuse. Many abstract concepts, such as a state machine, are
often reused across substantially different software products. However,
on the code level, their implementations are tangled with details
of particular products and often cannot be reused.

This loss of reuse can be attributed to the abstraction gap between
the high-level (concept level) and the low-level (code level) representations
of the solution. When programmers implement a high-level concept,
such as a state machine, they {}``compile'' the high-level concept
into code in a manual process. The product of this process is code
that integrates, often in an inseparable manner, the reusable knowledge
of how to code such a concept in the programming language in use (e.g.,
a state machine design patterns), with the specifics of the particular
instance of the concept (e.g., a particular instance of a state machine).

One solution to this problem is the use of \emph{Domain-Specific Languages
(DSLs)}. Programmers use DSLs to code high-level concepts directly.
The DSL implementation is responsible for specifying the meaning of
these concepts in terms of lower-level concepts. This can be done
either by compiling the DSL code into code in some pre-existing language,
or by interpreting it. Either way, the application code now consists
of two parts: the DSL code and the DSL implementation. The DSL code
conveys the specifics of the application, which is generally not reusable
but very concise. The DSL implementation conveys the knowledge of
expressing high-level concepts in terms of low-level ones, which is
often complicated, but highly reusable. This method thus allows us
to take reusable concepts and turn them into reusable code, expressed
as the DSL implementation.

Indeed, DSLs can be used to solve this abstraction gap and achieve
higher code reusability. However, for this method to take effect in
real-life software development, it has to be applied systematically
throughout the code. Real-life software is complex and diverse. It
usually uses many kinds of high-level concepts. Some are globally
relevant (e.g., a state machine), but some are only relevant to an
industry or a particular software product-line (SPL). 

Using DSLs for these concepts can allow reusing the logic behind them.
This means that DSLs must be developed for various aspects of the
software, and that these DSLs need to be able to interact, in the
places where one high-level concept touches another, e.g., when a
network event (one high-level concept) triggers a state transition
in a state machine (another high-level concept). Having such interactions
requires that the DSLs be implemented over some common platform that
allows DSLs to interact, both syntactically and semantically. This
approach to software development, which advocates the use of interoperable
DSLs to write software, is called \emph{Language Oriented Programming}
(\emph{LOP})~\cite{Ward:1994:LOP,Dmitriev:2005:LOP,Fowler:2005:LWK}.

The main challenge for realizing LOP in real-life software lies in
the need to develop and use DSLs. Here, the choice of techniques and
tools used for DSL implementation bears a great significance on the
practicality of LOP. For example, the traditional approach of using
standard compiler-generator tools such as Lex and Yacc or ANTLR to
implement DSLs can work properly for a pre-determined, limited set
of concepts, but will not allow DSLs to be defined as separate, reusable
but interoperable components. 

One important decision one needs to make is the choice between internal
and external DSLs~\cite{Fowler:2005:LWK}. \emph{External DSLs} are
DSLs implemented in form of a compiler, translator or interpreter
for the DSLs, while \emph{internal DSLs} (or \emph{embedded} \emph{DSLs}~\cite{Hudak:1996:BDE})
DSLs are {}``sub-languages'' defined from within a host language.
Internal and external DSLs have inherent trade-offs. On the one hand,
external DSLs provide more freedom in defining syntax and semantics,
but place the burden of implementing the language on the DSL developer.
On the other hand, internal DSLs are much easier to implement, as
they reuse most of the facilities provided by the host language, but
are constrained by its syntax and semantics. In addition, DSL interoperability
is supported naturally by internal DSLs (where all the DSLs are actually
code in the same host language), while interoperability is much harder
to achieve using external DSLs.

To date, two approaches have been presented to overcome these trade-offs,
namely \emph{language workbenches} and \emph{LOP languages}. Both
of these approaches allow to develop one kind of DSLs, while mitigating
its limitations relative to the other kind:
\begin{description}
\item [{Language~Workbenches}] Language workbenches are integrated development
environments (IDEs) for developing external DSLs. They ease the task
of defining and implementing DSLs by providing (meta) DSLs dedicated
for that task. They provide some tooling (auto-completion, definition
search, etc.) for the DSLs for free, or at very little cost, by leveraging
the DSL definition. Language workbenches, in contrast to other compiler-generation
tools, are made to support DSL interoperability. The most notable
language workbenches are MPS~\cite{Dmitriev:2005:LOP} and the Intentional
Domain Workbench~\cite{Simonyi:2006:IS}. They both use \emph{projectional
editing}, an approach were the program is a model edited through a
view, as a replacement for using text editing and parsing. This allows
syntactic integration of DSLs without causing ambiguity. With projectional
editing, disambiguation is done when entering the code, e.g., by selecting
the intended construct from a list or a menu.
\item [{LOP~Languages}] This is a new concept presented by our group~\cite{Rosenan:2010:DLP}.
These are programming languages oriented towards LOP, similarly to
how object-oriented programming languages are oriented towards OOP.
By our definition, LOP languages are made to host internal DSLs, while
providing two important features previously associated with language
workbenches and external DSLs. These are: projectional editing, and
the ability to define and enforce DSL schemata. The Cedalion language~\cite{Lorenz:2010:CLO}
is an example of such an LOP language, based on logic programming
for hosting internal DSLs, with a static type system to provide a
basic notion of DSL schema.
\end{description}
The main difference between these two approaches is in the relationships
between languages in each framework. In language workbench we can
identify three: the DSL code, the DSL implementation (the meta level),
and the workbench provided DSLs for implementing DSLs (the meta-meta
level). LOP languages, on the other hand, provide all these function
from within a single programming language. In a way, this is their
advantage, allowing reuse across these levels.

In this work we implemented twice, as a case study, a simple SPL of
\emph{calculator software}, using two LOP techniques. One of the implementations
is based on external DSLs and the other on internal DSLs. The differences
between the two implementations provides a comparison in terms of
the cost of reuse between external and internal DSL. It also provides
a deeper understanding of LOP and how LOP can generally address the
issue of code reuse in SPLs. 

Specifically, we implemented the complete SPL in MPS and another complete
implementation in Cedalion. We present the two implementations and
discuss the pros and cons of each method. The choice of MPS and Cedalion
as the implementation tools for this paper was made due the fact that
their main difference is in the choice of external (MPS) versus internal
(Cedalion) DSLs, thus providing a comparison between these two approaches.
In other LOP respects they are similar (projectional editing, DSL
schema).We concentrate on the cost of achieving code reuse in these
two approaches. We conclude that both approaches indeed support reusability
by providing easy-to-use DSLs that hide the complexity of translating
high-level concepts into low-level, executable ones. However, the
difference between these LOP approaches lies in the DSL implementation.
Implementing internal DSLs over a declarative language is easier and
more straightforward than implementing external DSLs over an imperative
language.

\section{Case Study: Calculator Product Line\label{sec:Case-Study:-Calculator}}

To get the feel of how practical and useful LOP can be, and to study
the implications of using internal versus external DSLs, we present
here a small comparative case study, where we use LOP to create a
tiny SPL for calculator software. Our measurements will be both qualitative
(how well did we manage to reuse code) and quantitative (the cost,
in terms of implementation time). We conduct this study using two
tools: the MPS language workbench, and the Cedalion LOP language.
\begin{description}
\item [{Meta-Programming\,System\,(MPS)}] This is a projectional language
workbench (i.e., a language workbench using projectional editing)
developed by Dmitriev and his team at Jetbrain's~\cite{Dmitriev:2005:LOP}.
It is mostly open source, and can be freely downloaded. This made
it a good candidate for this case study. Its website contains examples
and tutorials to help new users get up-to-speed. It features relatively
mature and very powerful projectional editing capabilities, overcoming
some of the usability problems traditionally associated with projectional
editing. DSL implementation is typically done by generating code in
a language called the {}``base language,'' which is, for all practical
purposes, Java. Implementing a DSL in MPS requires creating templates
and conversions for all DSL constructs into lower-level languages,
and eventually, into the base language.
\item [{Cedalion}] Cedalion is an LOP programming language, based on logic
programming. Logic programming provides a declarative way to define
DSL semantics, while its static type system provides a structural
definition (a schema) for the DSL. Like MPS, it features projectional
editing, which allows syntactic freedom for DSL developers, without
the danger of creating ambiguities, since disambiguation is done when
entering the code. Cedalion is open source (\url{http://cedalion.sourceforge.net}).
Its projectional editor is implemented as an Eclipse plug-in, using
a Prolog back-end. Cedalion, however, is a research tool developed
as a proof-of-concept and as such lacks the maturity that MPS provides.
Nevertheless, Cedalion is more than capable to implement the case
study at hand.
\end{description}

\subsection{The Problem Statement}

To examine the value of LOP for code reuse, and to compare between
internal and external DSLs for this purpose, we define a problem,
which we shall solve using the above tools. The problem statement
is as follows:
\begin{quote}
Develop an SPL of calculator software. All calculators have a key-pad
and a line-display. On the key-pad there are numerous keys for digits,
operators and functions. Pressing these keys simply append characters
to the line-display. There is also an {}``execute'' or {}``=''
button, which, when pressed, replaces the expression in the display
with either the number to which the expression evaluates to, or the
string {}``Syntax Error'', if the expression is invalid.
\end{quote}
Since we are interested in a SPL, we refer to a whole product-line
of such calculators. These calculators differ in their choice of operators,
functions, and even digits (e.g., a hexadecimal calculator), and how
they evaluate to numbers. Our goal in this case study would be to
try and reuse as much code as possible between different calculators
in this SPL.

\subsection{General Guidelines\label{sub:General-Guidlines}}

In this case study we focus on the part of the software that parses
and evaluates the string into a value, assuming the rest of the software
(e.g., the line editing) are inherently reusable between different
calculators.

We will implement these calculators using LOP. This means that we
will first identify the high-level concepts we need to describe \emph{a
calculator}, regardless of the specific instance (scientific, financial,
etc.). We then define a DSL to express these concepts formally, and
implement it. In this case study we ignore any pre-existing DSLs that
may address these concepts, since we would like to aim for the real-life
scenario where such DSLs are often unavailable or inapplicable for
various reasons. We then implement each calculator using the DSL we
developed. These implementations are expected to be concise and very
high-level, expressing the syntax of each particular calculator. All
the logic common across calculators is expressed in the DSL implementation.
Reuse of calculator features expected to be common to different calculators
(such as the parsing of numbers and basic arithmetic operations) is
beyond the scope of the case study, and will be addressed briefly
in Sections~\ref{sub:MPS-DSL-Code-Reuse} and~\ref{sub:CED-DSL-Code-Reuse}.

\section{SPL Implementation in MPS\label{sec:Implementation-in-MPS}}

We now describe the calculator SPL implementation in MPS. Due to space
limitation we keep the MPS-related implementation details as brief
as possible.

\subsection{Defining the DSL\label{sub:MPS-Defining-the-DSL}}

We begin by analyzing our calculator SPL, in order to figure out
what kind of DSL(s) we need to define for it. Our software needs to
do two things: (1) parse a string, according to some grammar; and
(2) calculate a numeric value based on that parsing. We therefore
wish to implement our calculator using a DSL that combines a grammar
(context-free) and the evaluation of expressions. This is somewhat
similar to an attribute grammar, where each production rule is associated
with a single value. Existing DSLs, such as Yacc~\cite{Johnson:1975:YYA}
can be considered here. However, as stated in Section~\ref{sub:General-Guidlines},
for the purpose of the case study we ignore pre-existing DSLs and
implement the ones we need. For the purpose of this discussion we
consider the '+' operator. Its syntax can be defined as:\begin{equation}
\begin{array}{rcl}
expr & \textrm{::=} & expr,\;'+',\; multExpr\end{array}\label{eq:LR-plus}\end{equation}
We would evaluate $expr$ for Eq.~\ref{eq:LR-plus} by summing the
values of the derived $expr$ and $multExpr$ non-terminals. This
could be formulated as:\begin{equation}
\begin{array}{rcl}
expr & \textrm{::=} & a=expr,\;'+',\; b=multExpr\;\{a+b\}\end{array}\label{eq:LR-plus-calc}\end{equation}
by binding the result of evaluating both arguments with variables
$a$ and $b$ (using the $=$ operator), and then specifying that
the entire phrase evaluates to $a+b$, inside the curly braces.

This notation is clear and concise, however, making it executable
is far from trivial. The grammar in Eq.~\ref{eq:LR-plus} has a head
recursion, making it non-LL (this is actually an LR grammar). Parsing
LR grammars is significantly harder than parsing LL grammars. LL grammars
can be parsed using recursive descent, with reasonable effort. Generating
a parser for even a subclass of LR (such as LALR(1)) is a much harder
task~\cite{Johnson:1975:YYA}. We therefore would like to restrict
ourselves to LL grammars, and for that we need to avoid head recursion.
To make Eq.~\ref{eq:LR-plus} an LL grammar, we need to replace the
head recursion with a tail recursion:\begin{equation}
\begin{array}{rcl}
expr & \textrm{::=} & \mathit{multExpr},\;\mathit{exprSuffix}\\
\mathit{exprSuffix} & \textrm{::=} & '+',\; expr\end{array}\label{eq:LL-plus}\end{equation}
This changes the way we calculate the value. We need to adopt a top-down
approach for the evaluation. Such calculation can be formalized as
follows: \begin{equation}
\begin{array}{rcl}
expr & \textrm{::=} & a=\mathit{multExpr},\; s=\mathit{exprSuffix}(a)\;\left\{ s\right\} \\
\mathit{exprSuffix}(a) & \textrm{::=} & '+',\; b=expr\;\left\{ a+b\right\} \end{array}\label{eq:notation}\end{equation}
An \emph{$expr$} consists of a prefix ($\mathit{multExpr}$) and
a suffix ($\mathit{exprSuffix}$). We first parse the prefix and bind
its value to variable~$a$. Then we parse the suffix, providing it
the value of $a$ as argument. The suffix modifies the value by adding
the right-hand value (variable~$b$) to the parameter~$a$. Finally,
$expr$ returns the value returned from the suffix.

The notation used in the example in Eq.~\ref{eq:notation} is sufficient
for expressing the logic of an entire calculator in our case study.

\paragraph{\textbf{DSL Schema}}

Now that we understand what our DSL looks like, we need to break it
down and understanding which constructs our DSL has, and more importantly,
how they are classified. The notation in Eq.~\ref{eq:notation} holds
four {}``families'' of constructs: Rules, Patterns, Reducibles and
Expressions. Most important is the distinction between patterns and
reducibles. Both patterns and reducibles define languages of strings,
however, a reducible reduces a string to a single value, whereas a
pattern reduces a string into a set of variable bindings. For example,
$'+',e=expr$ is a pattern, as it produces the bindings for $e$,
while the more complete term $'+',e=expr\left\{ p+e\right\} $ is
reducible, since it defines a single value ($p+e$) for the string
being parsed.

DSLs in MPS can rely on other languages. In this case, we use the
\emph{Expression} concept defined in the MPS \emph{base-language}~\cite{Dmitriev:2005:LOP}
as our expression type, so our language will inherit the wealth of
expressions supported by the base language with no effort on our part.
We do, however, need to define two expression concepts of our own:
a reference to an argument (such as $p$ in the term $\left\{ p+e\right\} $
in Eq.~\ref{eq:notation}), and to a bound variable (such as $e$
in the term $\left\{ p+e\right\} $ in Eq.~\ref{eq:notation}). These
new expression concepts will integrate seamlessly into base-language
expression concepts such as the '+' expression.

\begin{figure}[t]
\begin{centering}
\begin{tabular}{cc}
\subfloat[\label{fig:Concept-definition-alt}Concept definition]{\includegraphics[scale=0.5]{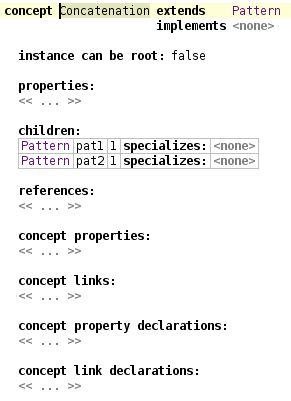}} & \subfloat[\label{fig:Editor-definition-alt}Editor definition]{\includegraphics[scale=0.5]{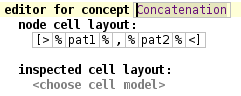}}\tabularnewline
\end{tabular}
\par\end{centering}

\caption{Definition of the Concatenation concept in MPS}

\end{figure}

In MPS, a DSL schema is defined by defining the language's \emph{structure}
\emph{model}. This model consists of \emph{concepts}, which are each
defined using its own form. The concept definition resembles a class
definition. It contains the concept's name, base-concept, implemented
interfaces, child concepts, referenced concepts, properties, etc.
For child and referenced concepts, cardinality should be provided.
Table~\ref{tab:List-of-concepts} lists the concept in our DSL. Figure~\ref{fig:Concept-definition-alt}
shows the definition of \emph{Concatenation}, as an example for a
concept definition. Note that this is a screenshot and not code listing,
due to MPS's projectional nature.%
\begin{table}[t]
\begin{tabular}{|>{\centering}p{1in}|>{\centering}p{0.6in}|c|>{\centering}p{2in}|}
\hline 
Concept & Base Concept & Projection & Description\tabularnewline
\hline
\hline 
Alternative & Reducible & $\begin{array}{lc}
a & |\\
b\end{array}$ & Choice between two reducibles\tabularnewline
\hline 
Concatenation & Pattern & $a,b$ & Concatenation of two patterns\tabularnewline
\hline 
Empty & Pattern & $<\textrm{empty}>$ & A pattern matching an empty string\tabularnewline
\hline 
Grammar & - & $\begin{array}{l}
\textrm{grammar}\, name\\
rules...\end{array}$ & A full grammar\tabularnewline
\hline 
NamedPattern & Pattern & $v=r$ & Assigning a name to the value produced by reducible $r$\tabularnewline
\hline 
NamedPattern Reference & Expression & $name$ & An expression evaluating to the value returned from parsing the reducible
associated with name\tabularnewline
\hline 
NonTerminal & Reducible & $name(args...)$ & References the rule named $name$, providing it arguments $args$\tabularnewline
\hline 
PatternValue & Reducible & $p\{e\}$ & Evaluates to the value of $e$, with the variable bindings received
from $p$\tabularnewline
\hline 
Rule & - & $name(args...)\textrm{::=}r$ & A production rule in the grammar\tabularnewline
\hline 
RuleArgReference & Expression & $name$ & An expression evaluating to the value of an argument given to the
rule\tabularnewline
\hline 
RuleArgument & - & $name$ & A formal argument for a rule\tabularnewline
\hline 
Terminal & Pattern & $'string'$ & A pattern matching a constant string\tabularnewline
\hline
\end{tabular}

\caption{\label{tab:List-of-concepts}List of concepts in the Grammar DSL}

\end{table}

\paragraph{\textbf{Defining the Editors}}

To allow projectional editing, we need to define how each concept
is visualized and edited. In MPS we do this by defining an \emph{editor
model}. Figure~\ref{fig:Editor-definition-alt} shows the editor
definition for the \emph{Concatenation} concept.

\paragraph{\textbf{Language Refinements}}

Now the language is defined, although we have not yet implemented
it. However, two refinements are in order:
\begin{enumerate}
\item Limiting the scope of rule arguments to the rule they are defined
in, and limiting the scope of variables to the pattern they are defined
in. These are done by defining a \emph{constraints model} for these
concepts.
\item Making the type of both variables and arguments {}``double,'' when
used in expressions. In addition, expressions associated with patterns
must also evaluate to {}``double.'' These rules are specified in
a \emph{type system model}.
\end{enumerate}
We omit screenshot of these definitions due to space limitations.

\subsection{Implementing the DSL}

A generator translates the DSL code into a lower-level, executable
language, making the DSL executable. This translation defines the
semantics of our DSL. Before implementing a generator we need to decide
on a target language. In MPS, if Java is an acceptable output language,
the MPS \emph{base-language}~\cite{Dmitriev:2005:LOP} will be a
natural choice. This is an adaptation of Java to MPS including most
of its features (MPS1.1 does not yet support generics), but like all
other MPS-based languages, it is edited using a projectional editor.

The more interesting question we need to ask is how do we wish to
see our DSL program translated to that target language (i.e., Java).
In our case, this means how do we wish to implement a parser or evaluator
in Java (or a Java-like language). We already mentioned that we prefer
top-down parsing (LL) over bottom-up (LR), since the latter requires
some heavy algorithms which we wish to avoid in this case. Therefore,
we need to understand how to implement a recursive descent parser
in Java. There are several ways to do that with performance--simplicity
trade-offs. Here we prefer simplicity over performance, and specifically
we prefer the simplicity of the \emph{generator}, and not necessarily
that of the \emph{generated code}.

The biggest challenge in this translation is the need for backtracking.
In this case, backtracking is used to support look-ahead. With backtracking,
the parser can go forward several characters following a certain alternative,
not find what it is looking for, and then backtrack to the point when
it made the choice and re-parse the text using a new alternative.
This technique is expected to be simpler (in terms of generator code)
then a possible alternative of turning the non-deterministic state
machine into a deterministic one, with no backtracking. One of the
main challenges of introducing backtracking is with regard to variable
bindings. In our DSL we bind values to variables. These values may
change due to backtracking. We need a way to save not only the state
of parsing, but also the value of variables, and restore them when
backtracking. Some declarative languages, such as Prolog, provide
natural support for backtracking. Variable bindings in these languages
obey backtracking. In fact, variables in these languages do not change
their value with time \emph{except} with backtracking.

The semantics of Java (and hence the MPS base-language) does not have
natural support for backtracking. Therefore, one of our challenges
would be to build backtracking {}``from scratch.''

\paragraph{\textbf{Implementing a Generator}}

\begin{figure}[t]
\centering{}\includegraphics[scale=0.4]{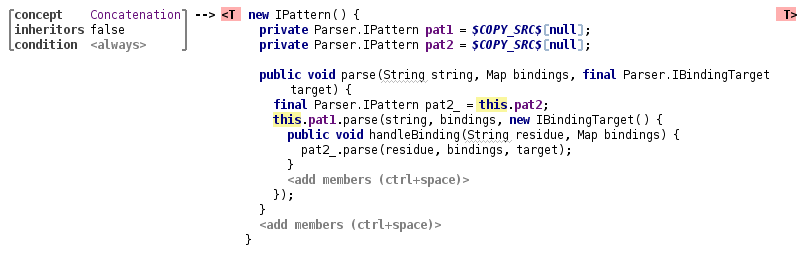}\caption{\label{fig:Reduction-rule-conc}Reduction rule for \emph{Concatenation}}

\end{figure}

Here we define the semantics of our DSL. This is done using \emph{mapping
rules} and \emph{reduction rules}. Mapping rules define how concepts
in the model map into top-level concepts in the generated code. A
class in the base-language is a top-level concept, so we map each
grammar to a class, using a mapping rule. The mapping rule specifies
a template of the class, which lays out the general structure of a
class generated to implement a grammar. This template uses macros
to customize the output class based on the properties and children
of the grammar. One kind of macro, \emph{COPY\_SRC}, is used to copy
child nodes into place in the template. This {}``copying'' includes
reduction where needed, following the reduction rules specified for
the generator. \emph{Reduction rules} define how a DSL concept is
translated to lower-level concepts, usually concepts of the base language.
In our DSL, reducibles and patterns have reduction rules, transforming
them into expressions in the base-language, resulting in an object
implementing \emph{IReducible} and \emph{IPattern} respectively. Figure~\ref{fig:Reduction-rule-conc}
shows the reduction rule associated with the \emph{Concatenation}
concept. It produces an instance that when getting a string it will
first pass it through the \emph{IPattern} associated with its left-hand
argument, passing each result (received using a callback) to the \emph{IPattern}
associated with its right-hand argument. The \emph{COPY\_SRC} macros
replace the \emph{null} values with the reduction of the left and
right-hand arguments of the concatenation.%

\subsection{Implementing the Calculator}

\begin{figure}[t]
\begin{centering}
\includegraphics[scale=0.4]{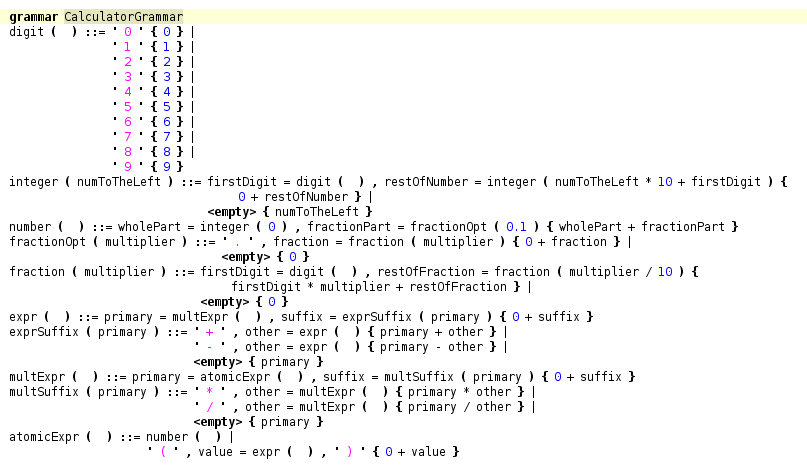}
\par\end{centering}

\caption{\label{fig:A-calculator-implementation}A calculator implementation}

\end{figure}

Now that our DSL is defined and implemented we can move forward to
using it to implement a concrete calculator. Figure~\ref{fig:A-calculator-implementation}
shows an implementation of a simple calculator, accepting numbers,
the four basic arithmetic operations and parentheses. This definition
is indeed short, concise, and contains nothing of the \emph{algorithm}
required to actually parse the string and to evaluate it. It only
contains the \emph{rules} by which this will be done.%

Each member of our product line should have such a definition, defining
its precise syntax and semantics. Since all implementation details
are encapsulated in the DSL definition (the generator model), they
are fully reused between these SPL instances.

\paragraph{\textbf{DSL Code Reuse}\label{sub:MPS-DSL-Code-Reuse}}

As concise as it may be, with complex enough calculators it may not
be enough to reuse the logic hidden in the DSL implementation. DSL
code duplication may become a problem as well. For example, the features
defined in Figure~\ref{fig:A-calculator-implementation} may be desired
in all calculators. Scientific calculators may add, e.g., trigonometric
functions, and financial calculators may add percentage calculations;
but both will keep this core behavior. One simple solution for that
would be to use inheritance, thus the scientific and financial calculator
grammars will inherit from the basic calculator grammar, adding their
own specific functionality. However, inheritance can go only a certain
way. Supporting an assortment of calculator, each with an arbitrary
selection of features will not work well with inheritance. Völter~\cite{Voelter:2010:IFV}
presents an approach to SPL engineering of DSL code in projectional
language workbenches, and has implemented it in MPS. With his approach,
DSL code can be annotated with feature-specific markers. A configuration
selecting the desired features controls code generation, so that only
the code that contributes to desired features takes effect. This approach
can be applied here, associating grammar rules with features. Consequently,
by enabling and disabling features we can control the insertion and
removal of grammar rules.

\section{SPL Implementation in Cedalion\label{sec:Implementation-in-Cedalion}}

\subsection{Defining and Implementing the DSL}

We wish to define and implement a DSL similar to the one described
in Section~\ref{sub:MPS-Defining-the-DSL}, but this time, we use
the internal DSL approach, where we implement each language construct
directly, and not by implementing a code generator for the language.
This difference allowed us to separate the language definition into
two separate DSLs: (1)~A {}``generic'' DSL for BNF grammars, and
(2)~an extension of that DSL to support evaluation ({}``Functional
BNF'', or FBNF). The concepts of \emph{Pattern} and \emph{Reducible}
exist here too, but the {}``generic'' BNF DSL only supports patterns,
while the FBNF DSL introduces reducibles. FBNF uses Cedalion's \emph{Functional}
DSL (a functional programming language over Cedalion) for expressions.
Table~\ref{tab:CED-List-of-concepts} shows all concepts in both
DSLs. There are only five of them (four in BNF and one in FBNF). This
is due to the fact that some concepts (e.g., variables, alternatives)
are inherent in Cedalion, due to its logic programming nature. Other
concepts, such as the $name(args...)$ reducible, will be defined
concretely for each reducible type, in the calculator definitions.%
\begin{table}[t]
\begin{tabular}{|c|c|c|>{\centering}p{2in}|}
\hline 
DSL & Concept & Type & Description\tabularnewline
\hline
\hline 
BNF & $A,B$ & pattern & Concatenation of two patterns\tabularnewline
\hline 
BNF & $\varepsilon$ & pattern & A pattern matching an empty string\tabularnewline
\hline 
BNF & $head\textrm{::=}body$ & statement & A production rule. Both $head$ and $body$ are of type \emph{pattern}.\tabularnewline
\hline 
BNF & $'string'$ & pattern & A pattern matching a constant string\tabularnewline
\hline 
FBNF & $Reducible\rightarrow^{Type}Expression$ & pattern & A pattern that associates a Reducible with an Expression of type Type.\tabularnewline
\hline
\end{tabular}

\caption{\label{tab:CED-List-of-concepts}List of concepts in the Cedalion
BNF DSL}

\end{table}
\begin{figure}[t]
\begin{centering}
\includegraphics[scale=0.3]{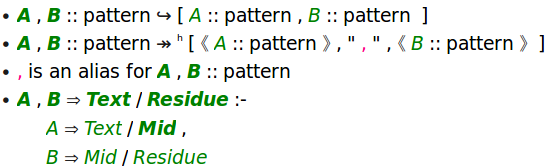}
\par\end{centering}

\caption{\label{fig:ced-conc}Implementation of the \emph{conc} concept in
Cedalion}

\end{figure}

Figure~\ref{fig:ced-conc} shows how a concept (in this case, $A,B$),
is defined and implemented in Cedalion. The first line is the type
signature (comparable with the \emph{concept definition} in MPS).
It defines $A,B$ to be a pattern, given that both $A$ and $B$ are
patterns. The second line is the projection definition, comparable
with MPS's \emph{editor definition}. It states that this concept shall
be displayed as a horizontal list (the tiny {}``h'') of visuals,
starting with a placeholder for the projection of $A$, followed by
a comma, followed by a placeholder for the projection of $B$. The
third line defines an alias for this concept, allowing the user to
type a comma and get auto-completion suggesting this concept. The
last line defines the semantics of $A,B$. It does so in a Prolog-like
manner, by contributing a clause to the $\mathit{Pattern}\Rightarrow\mathit{Text}/\mathit{Residue}$
predicate. This predicate states that $\mathit{Pattern}$ derives
a prefix $\mathit{Pref}$ of $\mathit{Text}$, such that $\mathit{Text}=\mathit{Pref}\cdot\mathit{Residue}$.
The clause here parses $\mathit{Text}$ as $A,B$ by first parsing
$\mathit{Text}$ as $A$, taking the residue $\mathit{Mid}$ and parsing
it as $B$. The residue now is the overall residue. Similar definitions
exist for all the other concepts. Backtracking and variable bindings
are handled implicitly, as they are inherent in logic programming,
simplifying the implementation significantly.

\subsection{Implementing the Calculator}

\begin{figure}[t]
\begin{centering}
\begin{tabular}{cc}
\multicolumn{2}{c}{\subfloat[\label{fig:CED-calc-imp}General expression syntax in Cedalion]{\begin{centering}
\includegraphics[scale=0.3]{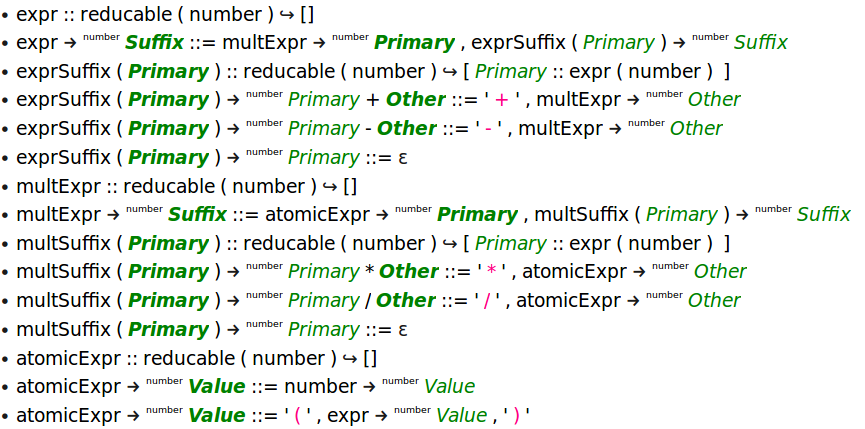}
\par\end{centering}

}}\tabularnewline
\subfloat[\label{fig:Trigonometric-functions}Trigonometric functions for scientific
calculators]{\begin{centering}
\includegraphics[scale=0.3]{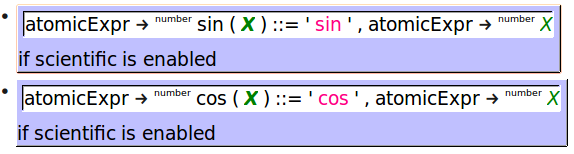}
\par\end{centering}

} & \subfloat[\label{fig:Configuration-example}Configuration example]{\begin{centering}
\includegraphics[scale=0.35]{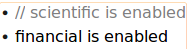}
\par\end{centering}

}\tabularnewline
\end{tabular}
\par\end{centering}

\caption{Calculator implementation in Cedalion}

\end{figure}

Figure~\ref{fig:CED-calc-imp} shows part of the implementation of
a simple calculator in Cedalion, using the BNF and FBNF DSLs we defined.
We omitted the part that defines the syntax of numbers, due to space
limitations. This definition is more elaborate then the one in Figure~\ref{fig:A-calculator-implementation}
due to the need to specify type signatures for all reducibles. Unlike
MPS, where concept definitions exist only in the DSL definition, in
Cedalion the DSL code is allowed and encouraged to define new concepts.
This allows safe usage of not only DSL constructs, but also of concepts
defined by the user, relieving the DSL developer from specifying custom
type system rules. While insisting on having type signatures present
in the code, Cedalion offers to add them automatically. The syntax
here is slightly different then the one we defined with MPS, because
while the DSL in MPS was designed as one monolithic DSL, here we see
a composition of two DSLs, trying to reuse their language constructs
as best we can. This is why we have the $Reducible\rightarrow^{Type}Expression$
concept on both sides of the production rules (on the right, replacing
the MPS \emph{NamedPattern} concept, and on the left, replacing the
\emph{PatternValue} concept (see Table~\ref{tab:List-of-concepts}).
The \emph{Alternative} in the MPS implementation is not needed here,
as different statements (or in this case, production rules), are taken
as having an \emph{or} relation, due to the nature of logic programming.

\paragraph{\textbf{DSL Code Reuse}\label{sub:CED-DSL-Code-Reuse}}

As in Section~\ref{sub:MPS-DSL-Code-Reuse}, two approaches can be
considered here: grammar inheritance or associating rules with features.
Since our BNF DSL does not have a concept of a grammar, the first
option is inapplicable (recall that this option has significant drawbacks).
However, associating rules with features is easy, and can be done
from outside the DSL~\cite{Voelter:2010:IFV}. Even though only full
statement can be associated with features, with feature variability~\cite{Voelter:2010:IFV}
this is not a limitation here, because we only intend to do so with
full production rules, which are statements. Figure~\ref{fig:Trigonometric-functions}
shows how do we support trigonometric functions only if the \emph{scientific}
feature is enabled. Figure~\ref{fig:Configuration-example} shows
a configuration, where the \emph{financial} feature is enabled, but
the \emph{scientific} feature is not.

\section{Results, Discussion and Related Work\label{sec:Results-and-Discussion}}

In previous sections we described a case study, where we used two
different tools: MPS and Cedalion, representing two different approaches
to DSLs, external using imperative base languages and internal using
a declarative host language, to construct a SPL of calculator software,
to achieve the goal of maximum code reuse between products. Indeed,
the use of DSLs (regardless of their implementation approach) improved
reusability by placing the complexity in a shared asset, the DSL implementation.
The particular assets in both implementations are stated in a high-level
language, capturing the high-level concepts of the problem domain.
With methods for associating DSL code with specific features, we can
maximize code reuse even at the DSL level, bringing code duplication
to zero. We therefore can conclude that we have achieved our goal
of code reuse through LOP.

But at what cost? Here the choice of tools takes effect. We measured
the time it took to implement and test the first, simplest calculator
(four arithmetic operations and parentheses), including the time it
took to define and implement the DSL behind it. With MPS it took us
about eight hours of work, most of which were dedicated to creating
the generator, which was not trivial (implementing backtracking and
variable bindings that adhere to backtracking in a Java-like language).
In Cedalion it took about two hours. The main challenge there was
dealing with the tool's sensitivity to user errors (i.e., its tendency
to crash due to them). As evidence for this difference in effort,
one can look at the complexity of the DSLs we defined in both tools.
It takes significantly less time to implement five constructs than
to implement twelve. Moreover, backtracking and variable binding were
given for free by the host language. No type system extensions were
needed, apart from defining a type signature for each construct. Once
the DSLs were defined and implemented, using them was relatively similar
in effort. MPS is more mature and therefore is more usable. Cedalion
requires type signatures for each new concept (including ones defined
in DSL code), which takes a little effort and makes the code a bit
more elaborate. However, these differences are minor relative to the
difference in effort in implementing DSLs. We therefore conclude that
from the view point of this case study, internal DSLs seam to be a
more cost effective for achieving code reuse through LOP.

\subsection{Threats to Validity}

In this work we used implementation time to measure cost efficiency.
It may be argued that our familiarity with Cedalion introduced a bias
in its favor. However, we took that into account, and familiarized
ourselves with MPS well enough before starting this case study, so
that the eight hours the implementation took did not include any of
the {}``learning curve.''

Another concern that may rise is the fact that we defined the case
study ourselves, and it may therefore be biased in favor of internal
DSLs, and Cedalion in particular. Specifically, the need for backtracking
and variable bindings turns the tables in favor of Cedalion. However,
these concepts are needed for many declarative notations. This is
why they are so fundamental in logic programming. We chose this case
study because it is relatively small and self contained, and at the
same time not trivial.

\subsection{Related Work\label{sec:Related-Work}}

The first notable work on code reuse through systematic use of DSLs
was done by Neighbor~\cite{Neighbors:1984:DAC}. This work introduces
Draco, a generative DSL framework. Draco's limitation in comparison
with MPS and Cedalion is in its dependence on parsing, which is sensitive
to conflicts that can arise when fusing the syntax of several DSLs
together.

The term LOP has been coined by Ward~\cite{Ward:1994:LOP}, who mentioned
reuse as one of its primary goals. It was then used by Dmitriev~\cite{Dmitriev:2005:LOP}
and Fowler~\cite{Fowler:2005:LWK}. Their notion of LOP is a bit
different than Ward's, as they emphasis the need for DSL interoperability.
DSL interoperability widens the opportunities for code reuse as the
DSLs become small, reusable components. However, Dmitriev~\cite{Dmitriev:2005:LOP}
and Fowler~\cite{Fowler:2005:LWK} do not explicitly mention code
reuse as a goal for LOP.

At the heart of this paper is a comparison of two approaches to LOP:
internal and external DSLs. To our knowledge, not many such comparisons
have been proposed. The Language Workbench Competition (LWC)~\cite{Voelter:2010:LWC}
provides a suggestion for comparison between language workbenches.
It provides a common task that should be implemented on different
workbenches to allow learning about their trade-offs. However, this
task does not tell a full story. It specifies a particular DSL, but
does not specify the semantics for that DSL. As a result, we found
the LWC less helpful for assessing reuse, and therefore turned to define
our own.

\section{Conclusion\label{sec:Conclusion}}

In this paper, we demonstrated how LOP can be used for code reuse,
allowing a separation-of-concerns between the generic, reusable high-level
concepts used to describe the problem and its solution, and the concrete
definition of a particular instance in a SPL. We showed that by defining
DSLs to capture high-level concepts we hide the complexity of transforming
them into low-level concepts inside the DSL implementation. The DSL
implementation becomes an asset shared across the SPL.

This LOP goal was achieved regardless of the choice of approach, internal
DSLs over a declarative language or external DSLs over an imperative
language. However, the cost of doing that differs significantly. In
our case study, using internal DSLs proved to be nearly four times
more cost-effective than using external DSLs. While the numbers may
vary based on the nature of the SPL and the ratio between the size
of the DSL implementations and the amount of DSL code, the advantage
of using internal DSLs is evident.

From a reuse perspective, internal DSLs provide an additional advantage.
Our ability to construct our DSL from two different DSLs (BNF and
FBNF) in the Cedalion implementation opens opportunities for reuse,
since the BNF DSL can be used by itself, possibly for totally different
kinds of products, and in conjunction with other DSLs. With MPS and
external DSLs, combining DSLs is also possible, however, because of
the code generation nature of the tool, we could not support such
a separation in our case study. We actually started with a generic
BNF DSL, but found it inapplicable for our needs, since it did not
support variable bindings.

The case study in this paper provides the reader unfamiliar with LOP
with a sense of how LOP can be leveraged for code reuse, and how language
workbenches and LOP languages can help performing that task. Our case
study shows an advantage for using declarative over the use of imperative
programming as a base language. Surprisingly, despite this demonstrated
(dis)advantage, the current state of the art is implementing LOP mainly
using imperative languages (through language workbenches), instead
of using declarative languages such as Cedalion.

\paragraph*{Acknowledgement }

We thank Micha\l{} \'{S}mia\l{}ek for his helpful comments.

\bibliographystyle{abbrv}

\end{document}